\documentclass[a4paper, aps, prb, twocolumn, showpacs, superscriptaddress]{revtex4}

\usepackage{amssymb}
\usepackage{amsmath}
\usepackage{color}
\usepackage{graphics, graphicx}
\usepackage{bbold}
\usepackage{psfrag}
\usepackage{mathcomp}

\newcommand{\IM}{\mbox{Im}}

\begin{document}

\author{D. Semmler} 
\affiliation{Institut f\"ur Theoretische Physik, Johann Wolfgang Goethe-Universit\"at, 60438 Frankfurt am Main, Germany}
\author{J. Wernsdorfer}
\affiliation{Institut f\"ur Theoretische Physik, Johann Wolfgang Goethe-Universit\"at, 60438 Frankfurt am Main, Germany} 
\author{U. Bissbort} 
\affiliation{Institut f\"ur Theoretische Physik, Johann Wolfgang Goethe-Universit\"at, 60438 Frankfurt am Main, Germany}
\author{K. Byczuk}
\affiliation{Institute of Theoretical Physics, Warsaw University, ul. Ho\.{z}a 69, 00-681 Warszawa, Poland}
\affiliation{Theoretical Physics III, Center for Electronic Correlations and Magnetism, Institute for Physics, University of Augsburg, 86135 Augsburg, Germany}
\author{W. Hofstetter}
\affiliation{Institut f\"ur Theoretische Physik, Johann Wolfgang Goethe-Universit\"at, 60438 Frankfurt am Main, Germany}

\date{\today}

\title{Localization of correlated fermions in optical lattices with speckle disorder}

\begin{abstract}
Strongly correlated fermions in three- and two-dimensional optical lattices with experimentally realistic speckle disorder are investigated. We extend and apply the statistical dynamical mean-field theory, which treats local correlations non-perturbatively, to incorporate on-site and hopping-type randomness on equal footing. Localization due to disorder is detected via the probability distribution function of the local density of states.  We obtain a complete paramagnetic ground state phase diagram for experimentally realistic parameters and find a strong suppression of the correlation-induced metal insulator transition due to disorder. Our results indicate that the Anderson-Mott and the Mott insulator are not continuously connected due to the specific character of speckle disorder. Furthermore, we discuss the effect of finite temperature on the single-particle spectral function.
\end{abstract}

\maketitle

\section{Introduction}

Anderson localization\cite{Anderson58,Kramer93} as well as strong correlation effects due to interactions\cite{Mott90,Imada98} have been studied intensely for decades in solid state physics. The elaborate interplay between these two fundamental phenomena gives rise to challenging questions\cite{Lee85,Altshuler85,Belitz94,Miranda05} for both experiment and theory. Recently, experiments with ultracold atoms in disordered lattices have opened an alternative route of investigation to gain a better understanding of this fundamental interplay.\cite{Aspect09,Sanchez-Palencia10} A high degree of tunability of the disorder and interaction strength is inherent to these experiments. 

Three different ways of realizing disordered optical lattices have been proposed: i) by loading an additional atomic species into the lattice and suppressing its mobility.\cite{Guenter06,Ospelkaus06} Thereby a binary disorder potential is realized due to the interatomic interaction; ii) by superimposing two laser beams with incommensurate frequencies,\cite{Damski03,Fallani07} thereby generating a stationary quasi-periodic lattice; iii) by using an optical speckle field\cite{Billy08} superimposed onto an optical lattice.\cite{White09,Pasiewski10}  

In general, particles in a disordered lattice are exposed to an on-site potential varying from site to site, which already gives rise to localization due to coherent backscattering.\cite{Anderson58} In tight-binding lattice models this randomness is termed diagonal disorder.  Additionally in a tight-binding description, the hopping amplitude between two lattice sites depends on the specific disorder potential as well, giving rise to off-diagonal disorder.\cite{Antoniou77,Soukoulis82,Ziman82,Dobros93,Inui94,Biswas00} A complete description of particles in a disordered lattice should incorporate both diagonal and off-diagonal disorder, i.e. random on-site energies and hopping disorder already in a single particle picture. 

A realistic theoretical description of ultracold atoms in a speckle disordered optical lattice should be capable of describing interaction and disorder on equal footing, preferably non-perturbatively. Furthermore, it should incorporate diagonal and off-diagonal disorder and, last but not least, should be able to treat other experimental features, such as finite temperature and the presence of the harmonic trap. 
In the case of bosons, there have been several theoretical works\cite{Bissbort09,Kruger09} combining some of these requirements. To the best of our knowledge, comparable studies have so far not been performed for fermions.

The aim of our paper is to close this gap and to provide a realistic description of strongly correlated fermions in a disordered optical lattice. For this purpose, we employ the statistical dynamical mean-field theory (DMFT)\cite{Dobros97,Song08,Miranda05,Semmler10} to solve the Anderson-Hubbard Hamiltonian numerically. The statistical DMFT incorporates both strong correlations and disorder-induced fluctuations and is known to give accurate results for high-dimensional systems, i.e. $d\geq 3$. We extend the statistical DMFT scheme to take hopping disorder into account. Realistic distributions for the Hubbard parameters, calculated by S.~Q.~Zhou and D.~M.~Ceperley,\cite{Zhou10} are used to describe the speckle disorder potential.

As our main result we find that the paramagnetic ground state phase diagram differs strongly from the phase diagram for pure diagonal box (homogeneous) disorder, as obtained within typical medium theory.\cite{Byczuk05,Byczuk10,Aguiar2009,Dobros10} In particular, the Mott insulator and the Anderson-Mott insulator are not continuously connected in  case of speckle disorder. Our results are reappraised by a complementary investigation of the two-dimensional square lattice by means of real-space DMFT (RDMFT).\cite{Dobros97,Snoek08,Helmes08}

\section{Model of correlated fermions in speckle disordered optical lattices}\label{system}
Ultracold fermionic atoms, such as $^6$Li or $^{40}$K, in disordered lattices are described by the Anderson-Hubbard Hamiltonian 
\begin{equation}
 H = -\sum\limits _{ i j \sigma } t_{ij} c_{i \sigma}^{\dagger} c_{j \sigma}   
 - \sum \limits_{i \sigma} (\mu - \epsilon_i) c_{i \sigma}^{\dagger} c_{i \sigma} +  \sum \limits_{i} U_i n_{i \uparrow } n_{i \downarrow }  \label{eq_sys1}
\end{equation}
where $c_{i \sigma}^{\dagger}$  ($c_{i \sigma}$) denotes the creation (annihilation) operator at a lattice site $i$ with spin $\sigma=\pm 1/2$. The fermionic number operator is given by $n_{i \sigma}= c_{i \sigma}^{\dagger}c_{i \sigma}$. The hopping amplitude between sites $i$ and $j$ is denoted by $t_{ij}$, the local interaction potential is parametrized by $U_i$, and the chemical potential is given by $\mu$. In the following, we consider fermions on a bipartite lattice with a semi-elliptical model density of states (DOS),\cite{Georges96,Eckstein,Kollar} which is characterized by the connectivity $K$. It is related to the lattice coordination number $Z$ via $K=Z-1$. The hopping amplitude $t_{ij}$ is assumed to be only non-zero between nearest neighbor sites $i$ and $j$. 

In experiments with cold atoms, the speckle disorder potential is created by a coherent laser beam that is scattered by a diffusor plate.\cite{Billy08,White09,Pasiewski10,Lye05,Clement06} A statistical analysis of the scattering process\cite{Goodman} shows that the probability distribution function (PDF) of the resulting light intensity pattern obeys $p_I(I)= \Theta(I) \,\exp(-I/\langle I \rangle)/ \langle I \rangle$, where $\langle I \rangle$ is the averaged light intensity and $\Theta(x)$ denotes the Heaviside function. By superimposing the speckle light pattern onto the optical lattice, the atoms are subjected to a random optical dipole potential $V_D(\mathbf{r}) \propto I(\mathbf{r})$,\cite{Clement06} which is attractive for red-detuned laser light or repulsive for blue-detuned laser light. We will consider the latter, i.e. a repulsive potential. Within the tight-binding model description this random potential gives rise to diagonal disorder, i.e. random on-site energies $\epsilon_i$, which are drawn from a PDF $p_{\epsilon}(\epsilon _i)$ given by 
\begin{equation}
p_{\epsilon}(\epsilon) = \frac{1}{\Delta} \exp(-\frac{\epsilon}{\Delta}) \Theta(\epsilon) \,, \label{PD_onsite_e}
\end{equation}
where $\Delta$ denotes the disorder strength. We note that this PDF of the on-site energies is unbounded from above, in contrast to other typically used distributions, such as box or binary disorder. Furthermore, we assume that the on-site energies of all lattice sites are independently and identically distributed. Within a tight binding model, the speckle disorder potential leads to off-diagonal disorder as well.\cite{White09,Zhou10} This so-called hopping disorder manifests itself in the random hopping amplitudes $t_{ij}$. For a given disorder distribution, inducing fluctuations in both the hopping and the on-site energies, the hopping coefficient $t_{ij}$ at a neighboring pair of sites is correlated with the difference in on-site energies $\Delta\epsilon=\epsilon_i-\epsilon_j$ and a realistic description requires modeling using a joint PDF\cite{Zhou10}
\begin{equation}
p_{\Delta\epsilon,t} (\Delta\epsilon,t) \neq  p_{\Delta\epsilon}(\Delta\epsilon) \cdot p_t(t) \,. 
\end{equation}
In our model (\ref{eq_sys1}), the joint PDF $p_{\Delta\epsilon,t}(\Delta\epsilon,t)$ is incorporated based on the data by S.~Q. Zhou and D.~M. Ceperley.\cite{Zhou10} We account for the dependence of these two random variables by a conditional PDF $p_t(t|\Delta \epsilon)$ for the hopping $t$ after the on-site energies on the respective sites were sampled. The details are given within the introduction of the method in subsection \ref{distributions} and the Appendix \ref{appendix}.

Furthermore, for a short-range interaction between particles the local Hubbard interaction parameter is proportional to the integral over the fourth power of the Wannier function on the same site, which in turn depends on the random lattice potential. Hence, the on-site interaction coefficient $U$ is a random variable as well, and the joint PDF $p_{\epsilon,U}(\epsilon,U)$ of the on-site interaction and the on-site energies also needs to be taken into account. 

\begin{figure}[tb]
\includegraphics[width=0.46\textwidth]{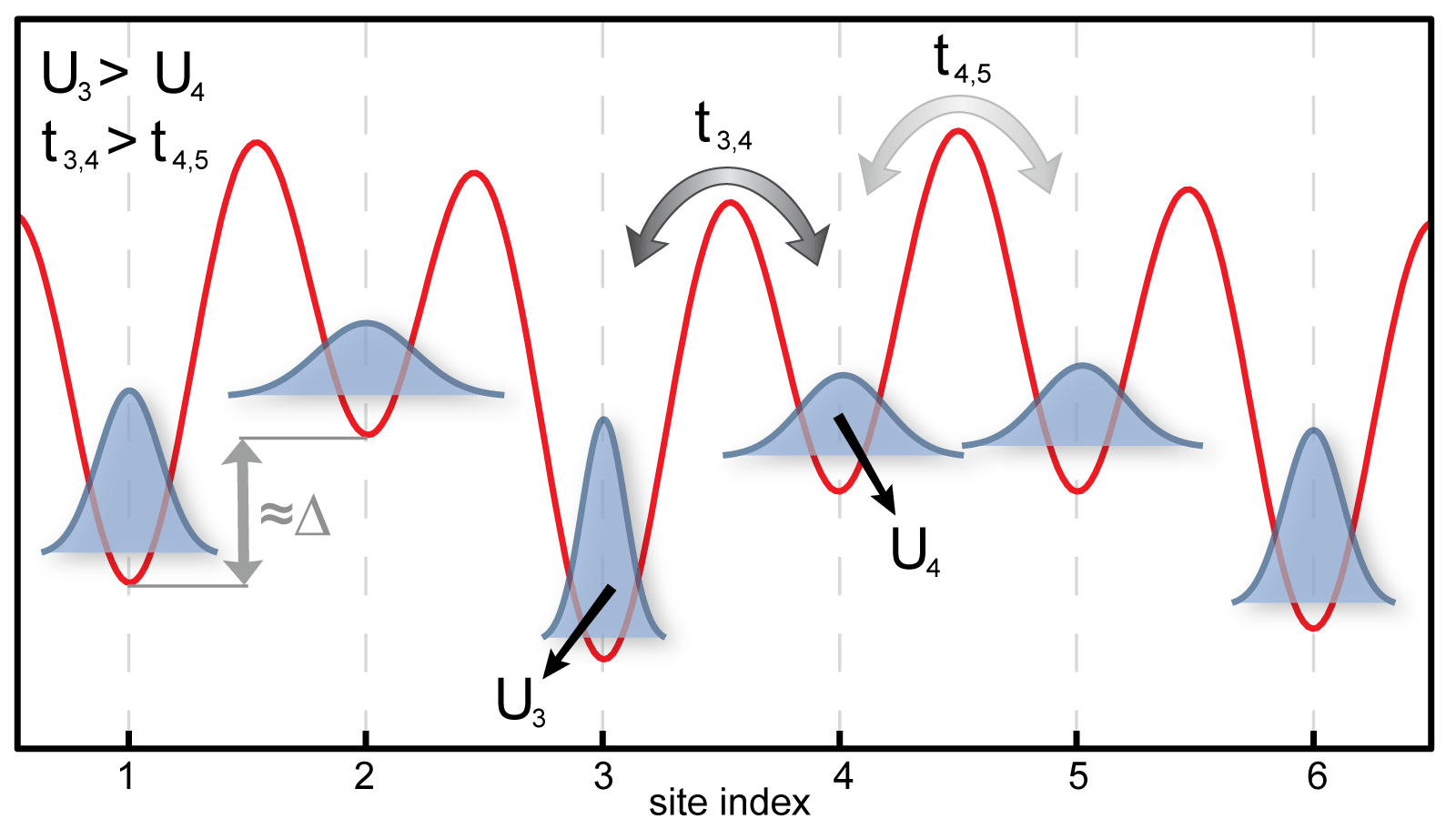}  
\caption{(Color online) One-dimensional illustration of the disordered lattice problem. The speckle field induces random on-site energies given by the probability distribution function $p_{\epsilon}(\epsilon)$. The hopping amplitudes $t_{ij}$ are random variables, which manifests itself in the probability distribution function $p_{\Delta\epsilon,t} (\Delta\epsilon,t)$. The local interaction strength $U_i$ depends on the local Wannier function, which results in a further probability distribution function $p_{\epsilon,U}(\epsilon,U)$.}
\label{fig_hubbard_parameters}
\end{figure}

Similar to the joint PDF of the difference in the on-site energies and the hopping coefficients, the joint PDF $p_{\epsilon,U}(\epsilon,U)$ is incorporated based on data from Ref. \onlinecite{Zhou10} and accounted for by the conditional PDF $p_U(U|\epsilon)$. The methodical details are explained in Sec. \ref{distributions} and Appendix \ref{appendix}. A one-dimensional illustration of the disordered lattice problem considered here is shown in Fig.~\ref{fig_hubbard_parameters}. 

%%%%%%%%%%%%%%%%%%%%%%%%%%%%%%%%%%%%%%%%%%%%%%%%%%
%%%%%%%%%%%%%%%%%%%%%%%%%%%%%%%%%%%%%%%%%%%%%%%%%%%

In experimental realizations, the disorder strength $\Delta$ is proportional to the speckle field strength $s_D=\langle V_D(\mathbf{r})\rangle$.\cite{Zhou10} and can therefore be tuned. The proportionality constant scales monotonically with the ratio of the speckle field autocorrelation length to the typical spatial extend of the Wannier function, and therefore depends on the experimental optical setup. In the experiment by White \textit{et al.},\cite{White09} the relation $\Delta=0.97s_D$ was found.\cite{Zhou10} The mean value of the interaction strength $U$ can be tuned by adjusting the $s$-wave scattering length $a$ between the two fermionic components by a Feshbach resonance\cite{Inouye1998,Theis04} or by varying the dimensionless lattice depth $s_L$, which in turn also influences the hopping amplitude. Therefore, keeping the lattice depth fixed, the disorder strength and the interaction strength can be tuned independently by varying the speckle field strength and the magnetic field. The relevant energy scale for the lattice depth $s_L$ and the speckle field strength $s_D$, is the recoil energy $E_R=\hbar k^2/2m$, where $k$ is the wave vector of the optical lattice and $m$ is the atomic mass. Throughout this paper we work in energy units of the noninteracting bandwidth $W_0=4 t \sqrt{K}$ of the homogeneous system, where the hopping amplitude $t$ can be related to the lattice depth $s_l$ by an exact band structure calculation, as for instance performed by M. Greiner.\cite{Greiner2003} The $s$-wave scattering length $a$ is given in units of the Bohr radius $a_0$.

\section{Methods for solving the model}\label{method}

Two conceptual approaches using DMFT to disordered systems, both referred to as \textit{statistical DMFT}, were originally introduced in Ref. \onlinecite{Dobros97}. In our opinion a distinction in terminology is justified and useful: Firstly, RDMFT, leading to a set of self-consistency equations for a fixed disorder realization or any inhomogeneity, thus constituting a deterministic approach. This method is applicable to any finite lattice structure and incorporates disorder or inhomogeneities non-perturbatively. 
Secondly, we refer to statistical DMFT as the intrinsically statistical approach to disordered systems on the infinite Bethe lattice. Here, the disordered system is investigated on a fully stochastic level and the self-consistency applies on a level of PDFs for the Green's function.
The statistical DMFT as well as the RDMFT are described in Secs. \ref{statDMFT} and \ref{RDMFT}. In Sec. \ref{distributions} we explain how we incorporate the joint PDFs of the Anderson-Hubbard model.

\subsection{Statistical dynamical mean-field theory}\label{statDMFT}

The statistical DMFT\cite{Dobros97} is a self-consistent computational scheme for determining the PDF of the local single-particle Green's functions, i.e. $p\left[G_{ii\sigma}(\omega)\right]$. Here, $G_{ij\sigma}(\omega)$ is the Fourier transform of the retarded Green's function $G_{ij\sigma}(t)=-i\Theta(t)\langle [ c_{i\sigma}(t),c_{j\sigma}^{\dagger}(0) ]_+ \rangle$, where $[..,..]_+$ denotes anticommutator brackets. In the following, only paramagnetic solutions of the Anderson-Hubbard model are considered. Therefore the spin index $\sigma$ is omitted hereafter. 

In the absence of interactions, the renormalized perturbation theory\cite{Economou_book} shows that the local Green's function can always be expressed as
\begin{equation}
G_{ii} (\omega) = \frac{1}{\omega + \mu -\epsilon_i - \Gamma_i(\omega) + i \eta} \,, \label{eq_M1}
\end{equation} 
where the hybridization function $\Gamma_i(\omega)$ describes the coupling of site $i$ to nearest neighbor lattice sites. For numerical reasons we introduce a finite broadening factor $\eta>0$. The limit $\eta\rightarrow0$ is important to detect localization. An additional self-energy $\Sigma_{ij}(\omega)$ accounts for interaction effects and is defined via a Dyson equation. In the dynamical mean-field theory\cite{Georges96} approximation the self-energy is local, i.e. $\Sigma_{ij}(\omega)\rightarrow \Sigma(\omega)\delta_{ij}$, and the local single-particle Green's function given by 
\begin{equation}
G_{ii} (\omega) = \frac{1}{\omega + \mu -\epsilon_i - \Sigma_i(\omega) - \Gamma_i(\omega) + i \eta} \,. \label{eq_M4}
\end{equation} 
The local approximation for the self-energy becomes exact in infinite dimensions, as was shown by W. Metzner and D. Vollhardt\cite{Metzner89} and is known to be a good approximation in three spatial dimensions. In a disordered system, translational invariance is broken and the local single-particle Green's function varies randomly from site to site, giving rise to a PDF $p\left[G_{ii}(\omega)\right]$. Within statistical DMFT, this PDF is determined by an ensemble with a large number $N$ of Green's functions. 

On the Bethe lattice, the hybridization function is given as sum over diagonal cavity Green's functions $G_{jj}^{(i)}(\omega)$,\cite{Abou73,Dobros97,Eckstein,Kollar}, i.e. 
\begin{equation}
\Gamma_i (\omega) = \sum\limits_{j=1}^{z} t_{ij}^2 G_{jj}^{(i)}(\omega).\label{eq_hybrid_cavity}
\end{equation} 
The corresponding cavity hybridization functions $\Gamma_j^{(i)}(\omega)$ are given as a sum over $z-1=K$ cavity Green's functions $G_{jj}^{(i,j)}(\omega)$. In this case, the exclusion of site $i$ is no longer relevant, since the Bethe lattice is loop-free. In a similar manner, subsequent equations reproduce the structural dependence of a sum over $K$ diagonal cavity Green's functions. 
An approximation first successfully applied by R. Abou-Chacra \textit{et al.},\cite{Abou73} and used in other works\cite{Abou74,Alvermann05} is that the cavity hybridization function is given by the hybridization function of the initial lattice
\begin{equation}
\Gamma_i (\omega) = \sum\limits_{j=1}^{K} t_{ij}^2 G_{jj}(\omega),\label{eq_hybrid} \;,
\end{equation}
with the sum now extending over $K$ Green's functions.   

The local self-energy $\Sigma_i(\omega)$ is determined by the solution of the corresponding impurity problem, defined by the given hybridization $\Gamma_{i}(\omega)$.\cite{Georges96,Dobros97} Therefore, the statistical DMFT maps the original lattice model onto an ensemble of impurities, whose PDF has to be determined self-consistently. 

%%%%%%%%%%%%%%%%%%%%%%%%%%%%%%%%%%%%%%%%%%%%%%%%%%
\begin{figure}[tb]
\includegraphics[width=0.32\textwidth]{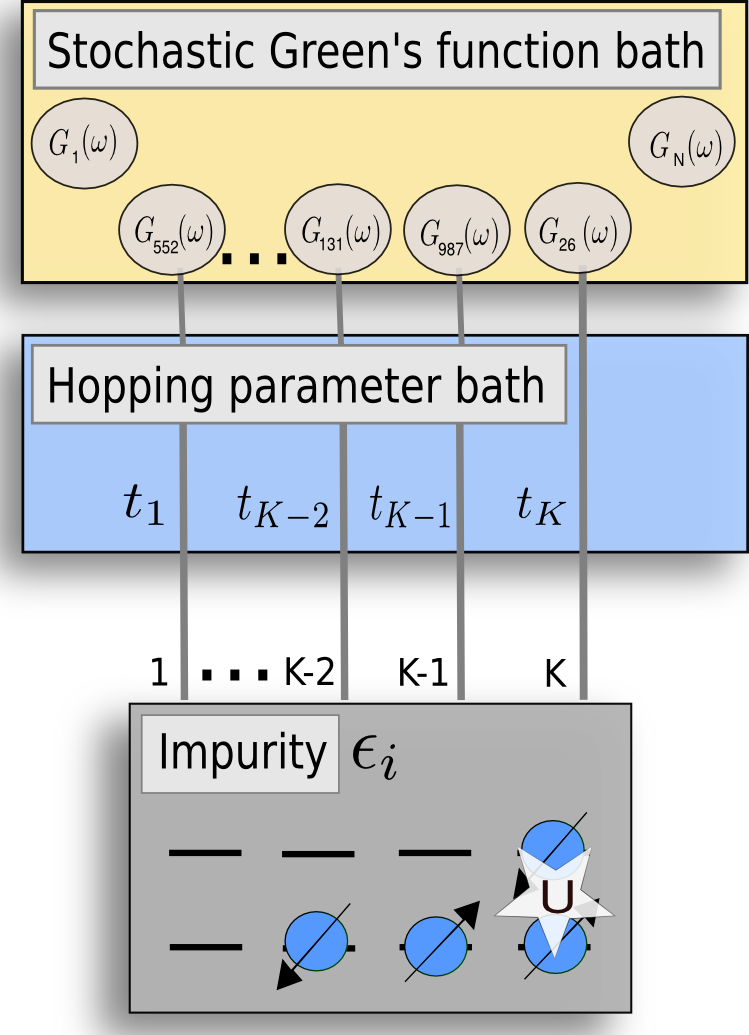}  
\caption{(Color online) Illustration of the statistical dynamical mean-field theory employed in this work. The many-body problem with disorder is mapped onto an ensemble of single impurities, which are coupled to a random ensemble of bath Green's functions, which is determined self-consistently. $G_n$ represents the $n$th sample from the ensemble of Green's functions. 
}
\label{figMethod}
\end{figure}
%%%%%%%%%%%%%%%%%%%%%%%%%%%%%%%%%%%%%%%%%%%%%%%%%%%

In practice, the self-consistent calculation scheme consists of the following steps: The starting point is an initial PDF $p\left[G_{ii}(\omega)\right]$ and the calculation is performed by (i)~drawing a random on-site energy $\epsilon_i$ from the PDF $p_{\epsilon}(\epsilon_i)$ given in Eq.~(\ref{PD_onsite_e}) and a random on-site interaction $U_i$ from the conditional PDF $p_U(U|\epsilon)$ for each sample of the ensemble; (ii) for each sample $i$ random hopping amplitudes $t_{ij}$ are drawn from the conditional PDF $p_t(t|\Delta\epsilon)$ depending on the difference $\Delta \epsilon=\epsilon_i-\epsilon_j$ of the on-site energies $\epsilon_i$ and $\epsilon_j$ of the randomly determined nearest neighbor $j$; (iii) The hybridization function $\Gamma_i(\omega)$, with the local single-particle Green's function $G_{jj}(\omega)$ of the nearest neighbors randomly sampled from the PDF $p\left[G_{ii}(\omega)\right]$, is determined via Eq.~(\ref{eq_hybrid}) for each sample; (iv) the local self-energy $\Sigma_i(\omega)$ is calculated from the solution of the local impurity problem by using an impurity solver; (v) the local single-particle Green's function $G_{ii}(\omega)$ is calculated using Eq.~(\ref{eq_M4}); (vi) having calculated a completely new ensemble $\{ G_{ii}(\omega)\}$, a new PDF $p\left[G_{ii}(\omega)\right]$ is obtained by construction of a histogram and we return to step (i). The algorithm is repeated until self-consistency for $p\left[ G_{ii}(\omega)\right]$ is achieved. We note that this method incorporates spatial fluctuations, i.e. quantum interference effects, caused by disorder via step (iii). Schematically the computational procedure is presented in Fig.~\ref{figMethod}.

The relevant physical observable is the local density of states (LDOS), given by $\rho_i (\omega) = - \frac{1}{\pi} \IM(G_{ii}(\omega))$, which is a random quantity in disordered systems. The corresponding distribution $p[\rho_i (\omega)]$ is obtained by counting all values of the LDOS for each frequency and by constructing a histogram. Statistical fluctuations are minimized by an artificial increase of the size of the ensemble after having reached self-consistency. This is done by constructing the histogram on the basis of typically $50-100$ successive DMFT iterations.
From this PDF the expectation value can be determined, i.e. the arithmetically averaged LDOS,
\begin{equation}
\langle \rho (\omega) \rangle_{\mbox{\tiny arith}} = \int  d\rho'(\omega) \; p[\rho'(\omega)] \, \rho'(\omega)
\end{equation}
and the typical value, which as in typical medium theory, is approximated by the geometrical average\cite{Dobros2003,Byczuk05,Byczuk09,Aguiar2009,Byczuk10,Dobros10} 
\begin{equation}
\langle \rho (\omega) \rangle_{\mbox{\tiny geom}} =  \exp \int d\rho'(\omega) \; p[\rho'(\omega)] \, \ln \rho'(\omega),
\end{equation}
where the dependence on $\omega$ is to be understood parametrically.
In the following, the cumulative PDFs 
\begin{equation}
P[\rho (\omega)] =  \int\limits_0^{\rho(\omega)} p[\rho'(\omega)] d\rho' (\omega) \,
\end{equation}
will also be useful to characterize the disordered system. 

We close the description of the method by a short discussion of our impurity solver. The most time-consuming part in the scheme is the solution of a large number of impurity problems, requiring a fast impurity solver. Here, we use iterative perturbation theory (IPT), \cite{MullerHartmann89,Potthof97,Kajueter97} which properly reproduces the non-interacting and atomic limits and was shown to qualitatively describe the Mott-Hubbard metal-insulator transition at a finite critical interaction strength $U_c$.\cite{Zhang93} Within IPT, the self-energy is calculated to the second order in $U$ in the non-renormalized perturbation expansion. Using this impurity solver, Green's function ensemble sizes of the order $N\sim10^3$ are computationally feasible within a parallelized code.

The original IPT, which was restricted to the half-filled case, was subsequently extended to densities away from half-filling. It is now commonly referred to as modified perturbation theory (MPT).\cite{Kajueter97,Potthof97} The self-energy within MPT is given by\cite{Potthof97}
\begin{equation}
\Sigma(\omega) = U n + \frac{a \Sigma^{(2)}(\omega)}{1 - b \Sigma^{(2)}(\omega)} \,, \label{eq_M6}
\end{equation}
where $\Sigma^{(2)}(\omega)$ is the second order perturbation contribution to the self-energy.\cite{MullerHartmann89} Within the perturbation expansion, the non-renormalized Hartree-Fock Green's functions\cite{Potthof97} 
\begin{equation}
G_i^{\mbox{\tiny{HF}}} (\omega) =  \frac{1}{\omega+\tilde{\mu}-\epsilon_i-U \langle n_i\rangle-\Gamma_i(\omega)+ i\eta}  
\end{equation}
are used as propagators. The parameter $\tilde{\mu}$ is fixed by requiring $\langle n_i \rangle = \langle n_i \rangle^{\mbox{\tiny{HF}}} $.\cite{Potthof97} The parameters $a$ and $b$ are determined such that the first three spectral moments 
\begin{equation}
M^{(m)} = \int\limits_{-\infty}^{\infty} \omega^m \rho (\omega) d\omega 
\end{equation}
with $m=0,1,2$ are reproduced and that the atomic limit is recovered correctly.\cite{Potthof97} For further details, the reader is referred to the work by M. Potthoff \textit{et al.}\cite{Potthof97} and our earlier work.\cite{Semmler10}

\subsection{Real-space dynamical mean-field theory}\label{RDMFT}

Fermions with a semi-elliptic DOS in many ways exhibit qualitatively the same physics as fermions in three-dimensional lattices. It is however important to note that qualitative differences may arise in lower dimensions. For this reason we also investigate fermions on a square lattice within RDMFT.\cite{Dobros97,Snoek08,Helmes08,Song08} Besides describing the Mott-Hubbard metal-insulator transition and magnetic order, RDMFT is also capable of treating spatial inhomogeneities such as disorder. As in DMFT, each lattice site is mapped onto a single-impurity Anderson Hamiltonian within RDMFT, where the hybridization function has to be determined self-consistently.

Starting with an arbitrary hybridization function, the solution of each impurity problem is provided by MPT, as in statistical DMFT, and leads to a set of on-site self-energy functions $\Sigma_{}^{i}\delta_{ij}(i\omega_n)$. They determine the self-energy matrix in the  real-space representation
\begin{equation}
(\mathbf{\Sigma})_{ij}=\Sigma_{}^{i}\delta_{ij}\,.
\end{equation}
Following the Dyson equation, the interacting lattice Green's function is given by
\begin{equation}\label{eq:int_gr}
 \mathbf{G}(i\omega_n)^{-1}=\mathbf{G}_{0}(i\omega_n)^{-1} - \mathbf{\Sigma}(i\omega_n)\,,
\end{equation}
where $\omega_n$ are the Matsubara frequencies. The non-interacting Green's function $\mathbf{G}_{0}(i\omega_n)$ in real-space representation is given by
\begin{equation}
 \mathbf{G}_{0}(i\omega_n)^{-1}=(\mu_{}+i\omega_n)\mathbf{1}-\mathbf{t}-\mathbf{V}\,,
\end{equation}
where $\mathbf{1}$ is the unity matrix, $\mathbf{t}$ is the matrix of hopping amplitudes, and $(\mathbf{V})_{ij}=\varepsilon_i\delta_{ij}$ denotes the matrix of random on-site energies.
Together with Eq.~(\ref{eq_M4}) and the diagonal elements from inverted relation~(\ref{eq:int_gr}) a set of local hybridization functions $\Gamma^{(i)}(\omega)$ is extracted, which closes the self-consistency loop.

%%%%%%%%%%%%%%%%%%%%%%%%%%%%%%%%%%%%%%%%%%%%%%%%%%
\begin{figure}[tb]
\includegraphics[width=0.45\textwidth]{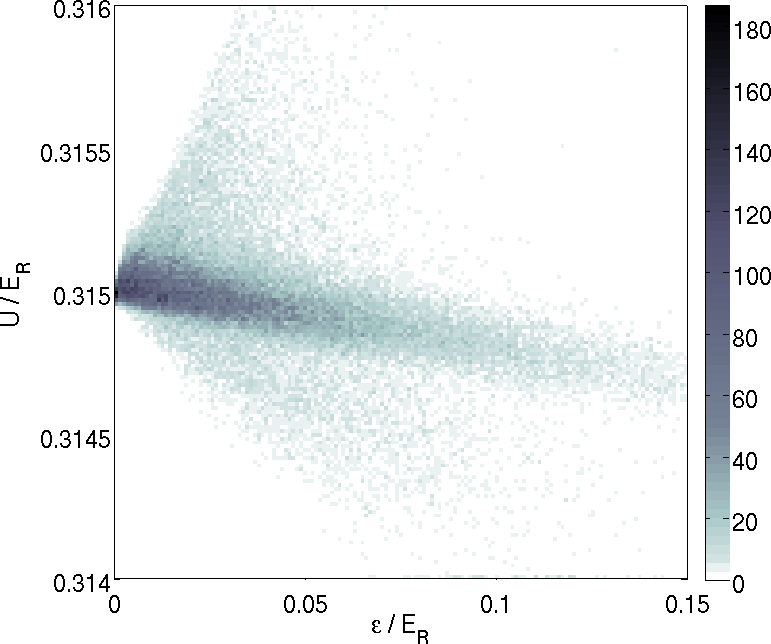}  
\caption{(Color online) Color coded probability distribution function (histogram) $p_{\epsilon,U}(\epsilon,U)$ of the on-site energies and the on-site interaction for speckle field strength $s_D=0.05E_R$ and $s$-wave scattering length $a=100 a_0$ for a 3d lattice. The lattice depth is given by $s_L = 10 E_R$.}
\label{fig_pd_e_U}
\end{figure}
%%%%%%%%%%%%%%%%%%%%%%%%%%%%%%%%%%%%%%%%%%%%%%%%%%%

\subsection{Joint probability distribution functions for the Hubbard parameters}\label{distributions}

The Hubbard parameters used both in the statistical DMFT and the real-space DMFT simulations are randomly generated based on the distributions obtained by S. Q. Zhou and D. M. Ceperley\cite{Zhou10} within an imaginary time evolution approach for the Wannier functions. Although these calculations were originally performed for bosonic $^{87}$Rb, the random parameters $\epsilon_i, U_i, t_{ij}$ depend only on the structure of the single particle states and can thus also be used for fermionic system after an appropriate rescaling.

The three sets of parameters for the Hubbard model underly statistical fluctuations. For the 3D case with equal laser intensity along each of the three axes, all parameters are unique functions of $s_L$ and $s_D$ for a given atomic species, when expressed in units of $E_R$. This  is not the case in the two-dimensional lattice, relevant for our real-space DMFT calculations. Here, the interaction parameter $U$ depends on the shape and strength of the axial trapping potential, which may vary significantly in different experiments. In this anisotropic case with fixed $a_s$, the lattice depth $s_L$ does not uniquely characterize the point in the phase diagram. Here we therefore give energies in units of the noninteracting  bandwidth $W_0$.

\begin{figure}[tb]
\includegraphics[width=0.45\textwidth]{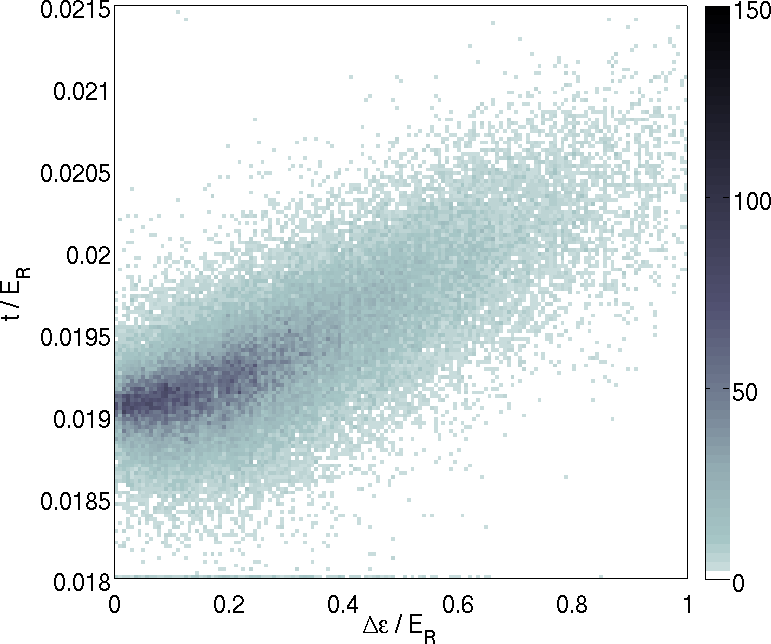}  
\caption{(Color online) Color coded joint probability distribution function (histogram) $p_{\Delta\epsilon,t}(\Delta\epsilon,t)$ of the difference in on-site energies between neighboring sites of a 3d lattice and the respective hopping amplitudes for a speckle field strength $s_D=0.4 E_R$ and lattice depth $s_L=10 E_R$. As restricted by the symmetry of $t_{ij}$ under the exchange of $\epsilon_i\leftrightarrow\epsilon_j$, the distribution only depends on $|\Delta\epsilon|$.}
\label{fig_distri_de_t}
\end{figure}

The on-site energies $\epsilon_i$ and interaction parameters $U_i$ are sampled from the data given in Ref.~\onlinecite{Zhou10} inherently containing  correlations, as shown in Fig.~\ref{fig_pd_e_U}. Up to a very good approximation, the on-site energies $\epsilon_i$ and tunneling parameters $t_{ij}$ are independent, however there is a significant correlation between the difference in on-site energies $\Delta \epsilon_{ij}=\epsilon_i-\epsilon_j$ and the tunneling parameter $t_{ij}$, as is shown in Fig.~\ref{fig_distri_de_t}. These correlations are taken into account both within the statistical DMFT and the RDMFT simulations by sampling $t$ from a conditional distribution $p(t| \Delta \epsilon)$ for a given (previously sampled) $\Delta \epsilon$. More technical details of how the conditional PDFs are constructed are given in Appendix \ref{appendix}.

%%%%%%%%%%%%%%%%%%%%%%%%%%%%%%%%%%%%%%%%%%%%%%%%%%
%%%%%%%%%%%%%%%%%%%%%%%%%%%%%%%%%%%%%%%%%%%%%%%%%%%

\section{Results}\label{results}

In our calculations we consider ultracold $^{40}$K atoms in a mixture of two hyperfine states in an optical lattice generated by a laser with a wavelength $\lambda_L=738\mbox{nm}$. The lattice depth is fixed to $s_L=10 E_R$. Moreover, we consider the half-filled case, i.e. band filling $\nu=\frac{1}{N}\sum_{i\sigma}\langle  n_{i\sigma}\rangle =1$, which is accomplished by adjusting the chemical potential $\mu$. For the statistical DMFT calculations, the lattice connectivity $K=6$ was chosen.

Describing the physics of strongly correlated fermions in a speckle-disordered optical lattice necessitates a proper definition of the relevant phases. The \emph{Mott insulator} is incompressible and its spectrum is gapped. Its correlation gap is proportional to the interaction strength $U$. A gapped spectrum in turn means that the arithmetic average of the LDOS at the Fermi level vanishes $\langle \rho(\omega=0)\rangle_{\mbox{\tiny arith}}=0$. A second insulating phase which we find here is the \emph{Anderson insulator}\cite{Anderson58} for zero interaction. This phase is compressible and its spectrum is point-like, which can be attributed to an absence of diffusion.\cite{Economou72} In the interacting case, we define a state to be localized or extended, if the spectrum of the single-particle Green's function is point-like or continuous respectively. The \emph{Anderson-Mott insulator} is defined by localized single-particle excitations at the Fermi level. Recently, it was shown that localized single particle excitations imply a vanishing conductivity in case of weak interactions.\cite{Basko2006} The detection of Anderson localization within statistical DMFT will be explained in the following subsection \ref{sec_localization}. Finally, the \emph{paramagnetic metal} is compressible and therefore has a non-vanishing LDOS $\langle \rho(\omega=0)\rangle_{\mbox{\tiny arith}}$ at the Fermi level, where the states are extended.

%%%%%%%%%%%%%%%%%%%%%%%%%%%%%%%%%%%%%%%%%%%%%%%%%%
\begin{figure}[tb]
\includegraphics[width=0.47\textwidth]{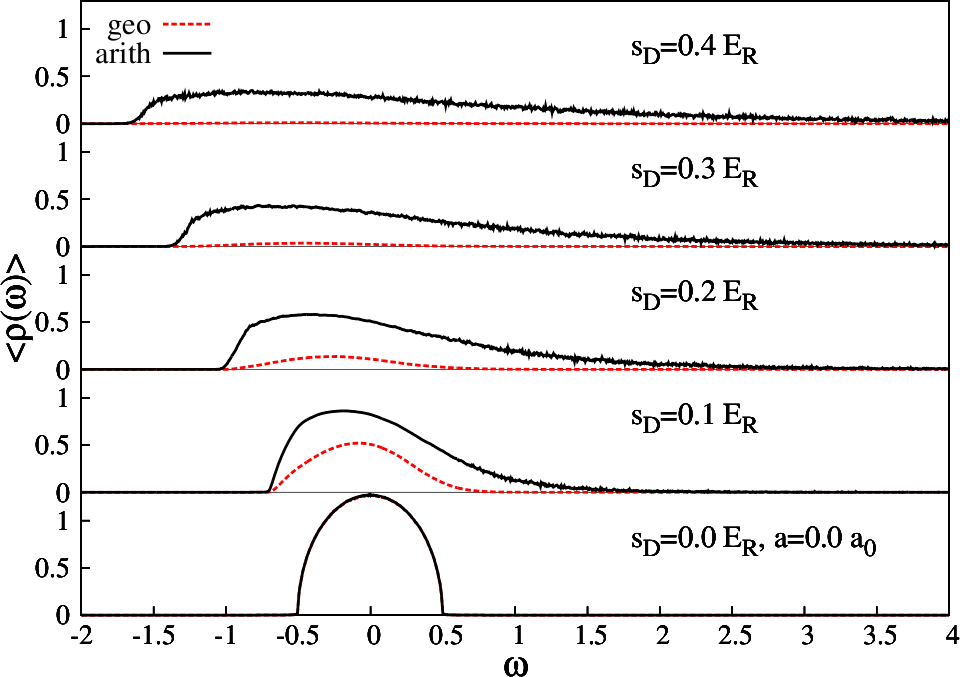}  
\caption{
(Color online) Comparison of arithmetically (black solid lines) and geometrically (red dashed lines) averaged spectral functions with increasing speckle field strength $s_D$ ($s_D=0.0E_R, 0.1E_R, 0.2E_R,0.3E_R,0.4E_R$) in the non-interacting limit $a=0$. Parameters are $\nu=1.0$, $s_L = 10 E_R$, and $\eta=10^{-3}$.}
\label{figArithGeoNOIA}
\end{figure}
%%%%%%%%%%%%%%%%%%%%%%%%%%%%%%%%%%%%%%%%%%%%%%%%%%%

\subsection{Detecting Anderson localization within statistical DMFT}\label{sec_localization}

%%%%%%%%%%%%%%%%%%%%%%%%%%%%%%%%%%%%%%%%%%%%%%%%%%
\begin{figure}[b]
\includegraphics[width=0.47\textwidth]{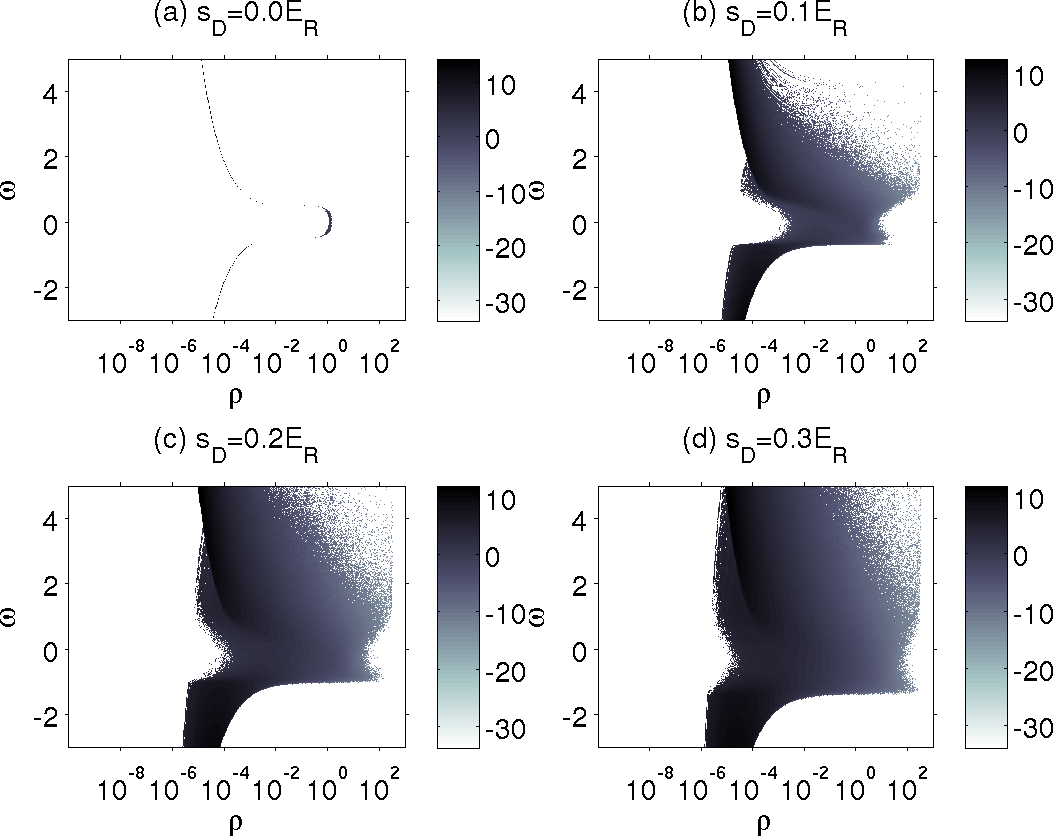}  
\caption{(Color online) Color coded natural logarithm of PDF $p[\rho]$ for increasing speckle field strength $s_D$: (a) $s_D=0.0 E_R$, (b) $s_D=0.1 E_R$, (c) $s_D=0.2 E_R$, and (d) $s_D=0.3 E_R$. Parameters are $a_s=0$, $\eta=10^{-3}$, $\nu=1.0$, and $s_L = 10 E_R$.}
\label{figsd}
\end{figure}
%%%%%%%%%%%%%%%%%%%%%%%%%%%%%%%%%%%%%%%%%%%%%%%%%%%  

The arithmetic and geometric averages of the LDOS obtained within statistical DMFT, which for the non-interacting case reduces to the local distribution approach,\cite{Abou73,Alvermann05} exhibit different behavior with increasing disorder strength $\Delta$. Fig.~\ref{figArithGeoNOIA} displays the evolution of the arithmetically and geometrically averaged LDOS in the non-interacting case, when the speckle field strength is increased from $0 E_R$ to $0.4 E_R$. First of all, we note that with an increase in the speckle field strength, the spectra are broadened and long tails emerge at positive energies. A noticeable difference between the geometric mean and the arithmetic mean, is that the geometric mean of the LDOS converges to zero with increasing $\Delta$. This can be attributed to the transition from extended states to localized states, which is also observed within typical medium theory.\cite{Dobros2003,Byczuk05,Byczuk09,Aguiar2009,Dobros10} However, not only averages, but also the full PDF $p[\rho(\omega)]$ is accessible within the statistical DMFT, which enables a more accurate detection of localization than attributing it to a vanishing geometrically averaged LDOS. This approach will be explained in detail in the next paragraph. The PDFs $p[\rho(\omega)]$ associated with the data shown in Fig.~\ref{figArithGeoNOIA} are plotted in Fig.~\ref{figsd}. As expected in the non-disordered case (cf. panel (a)), the PDF for each frequency is given by a delta function. For finite disorder strength, the PDFs extend over several orders of magnitude.  

%%%%%%%%%%%%%%%%%%%%%%%%%%%%%%%%%%%%%%%%%%%%%%%%%%
\begin{figure}[tb]
\includegraphics[width=0.47\textwidth]{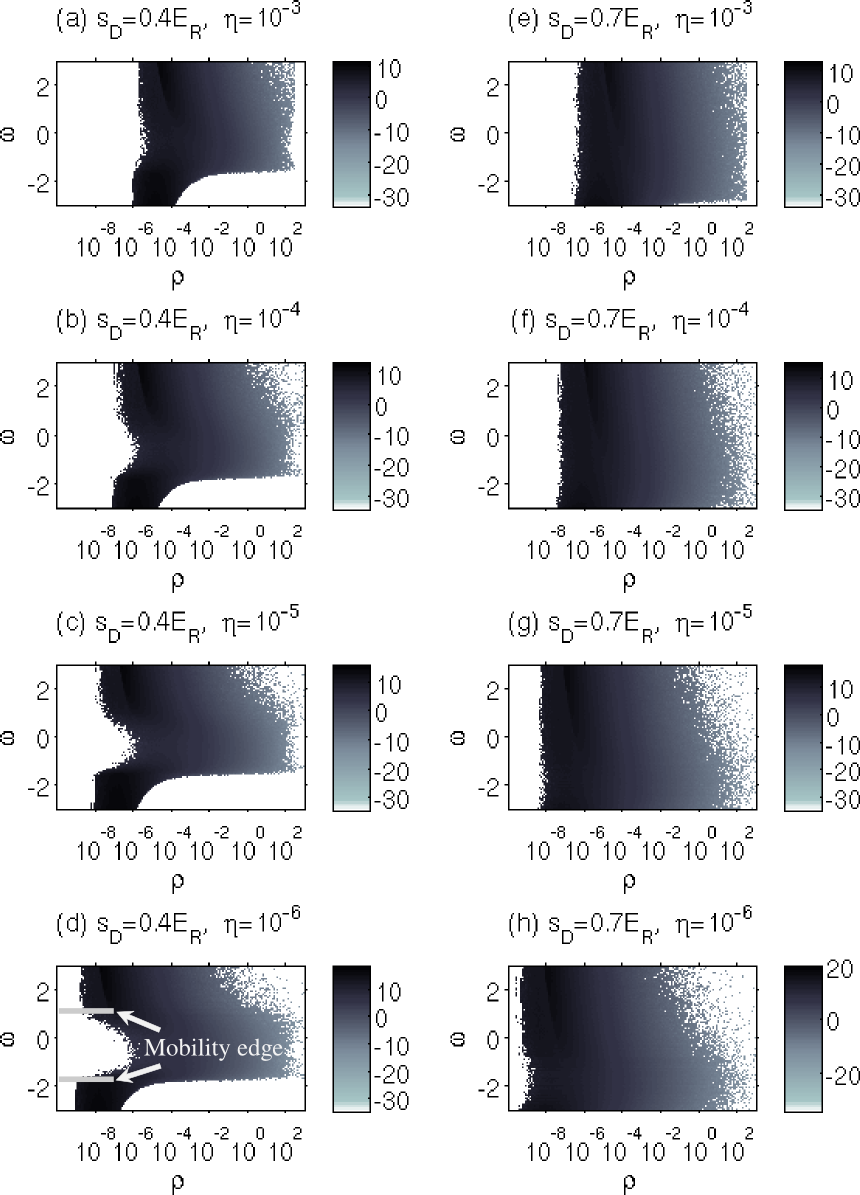}  
\caption{(Color online) Color coded natural logarithm of PDFs $p[\rho(\omega)]$ for decreasing broadening $\eta$ and speckle field strength $s_D=0.4 E_R$ (panels (a)-(d)) and  $s_D=0.7 E_R$ (panels (e)-(h)) of non-interacting fermions. Parameters are $\nu=1.0$ and $s_L = 10 E_R$.}
\label{fig_scale_sd0,4_sd0,7}
\end{figure}
%%%%%%%%%%%%%%%%%%%%%%%%%%%%%%%%%%%%%%%%%%%%%%%%%%%

Within statistical DMFT, extended and localized states are characterized by different behavior of the PDF $p[\rho(\omega)]$ in the limit of vanishing broadening $\eta\rightarrow 0$.\cite{Semmler10,Alvermann05} This procedure is motivated by the fact that the extended states are given by a branch cut on the real axis of the single-particle Green's function, whereas the localized states are characterized by a dense distribution of poles in the thermodynamic limit.\cite{Economou72} Let us consider a single particle initially
located at site $0$ on a finite lattice and let $|\psi_n\rangle$ denote the complete set of single particle energy eigenstates on the lattice with eigenenergies $E_n$.  The local Green’s function then takes on the form \cite{Licciardello1975}
\begin{equation}
G_{00}(\omega)= \sum\limits_n \frac{f_n}{\omega-E_n} \;. \label{eq_ef_representation_GF}
\end{equation}
Here, $f_n= \langle0|\psi_n \rangle \langle\psi_n|0 \rangle$ denotes the overlap of the eigenstate $n$ with the Wannier function on site $0$. The return probability is given by $p_{0\rightarrow 0}(t\rightarrow \infty)=\sum_n f_n^2$.\cite{Licciardello1975}  In case of an extended eigenstate, the residue $f_n$ is proportional to the inverse number of occupied lattice sites $N^{-1}$ and the return probability approaches zero in the limit of an infinite system. In contrast, for a localized state the spectrum is given by a dense distribution of poles in the allowed energy interval on the real $\omega$-axis. The residues approach a finite value, but some will dominate the sum in equation (\ref{eq_ef_representation_GF}). If their values were sorted by value, their contribution would decrease exponentially for spatially localized states. In particular, if we introduce a small coupling $\eta$ to a dissipative bath and consider a contour which encloses a small energy interval $\delta E$, then the most probable value of the sum of the residues of poles enclosed by the contour will decrease exponentially proportional to the ratio $\Delta/\delta E$. Accordingly, the most probable value of the imaginary part of the Green's function is proportional to $\eta$.\cite{Thouless1970} Hence, the LDOS will be highly fragmented in case of localized eigenstates, characterized by dominating well-separated resonances. In contrast, the arithmetically averaged spectral function, i.e. the DOS of the system, does not exhibit this high fragmentation, since the spectral weight must be located somewhere in the lattice for every energy. Consequently, the maximum of the PDF of the LDOS of an extended state saturates at a finite value for $\eta\rightarrow 0$, while an increasing amount of the PDF's weight shifts to zero in the case of localized states.

%%%%%%%%%%%%%%%%%%%%%%%%%%%%%%%%%%%%%%%%%%%%%%%%%%
\begin{figure}[tb]
\includegraphics[width=0.47\textwidth]{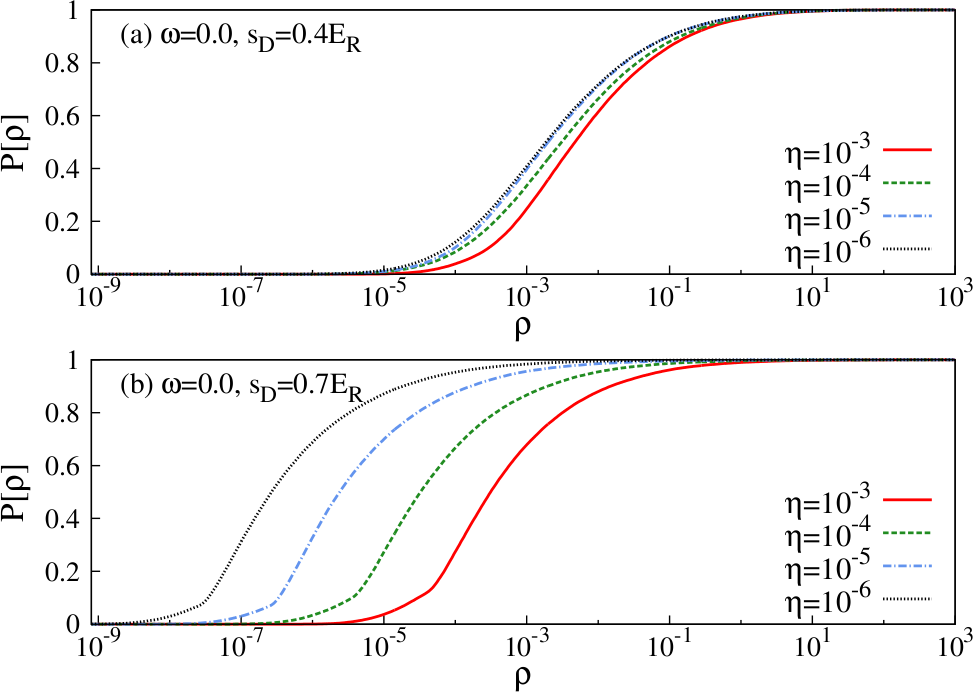}  
\caption{(Color online) Behavior of the cumulative PDFs $P[\rho(\omega=0)]$ in the noninteracting case as the broadening $\eta$ is decreased from $10^{-3}$ to $10^{-6}$ for (a) speckle field strength $s_D=0.4 E_R$ and (b) speckle field strength $s_D=0.7 E_R$. Parameters are $a=0$, $\nu=1.0$, and $s_L = 10 E_R$.}
\label{fig_scale_2d}
\end{figure}
%%%%%%%%%%%%%%%%%%%%%%%%%%%%%%%%%%%%%%%%%%%%%%%%%%%

In Fig.~\ref{fig_scale_sd0,4_sd0,7} the behavior of the PDFs at two speckle field strengths $s_D=0.4 E_R$ and $s_D=0.7 E_R$ is plotted for a sequence of decreasing broadening $\eta\rightarrow 0$. The different behavior of the extended and localized states allows us to distinguish between them and the mobility edges can be identified as shown in panel (d) for $s_D=0.4 E_R$. The states between the  mobility edges are extended, whereas the ones outside are localized. We observe that the states at the Fermi level are extended for speckle field strength $s_D=0.4$ which is therefore metallic, whereas for speckle field strength $s_D=0.7 E_R$ the states at the Fermi level are localized. This can also be seen from the 2d-plots of the corresponding cumulative PDFs at the Fermi level for the two speckle field strengths in Fig.~\ref{fig_scale_2d}.

\subsection{Paramagnetic ground state phase diagram of the interacting system within statistical DMFT}\label{sec_pd}

\begin{figure}[tb]
\includegraphics[width=0.47\textwidth]{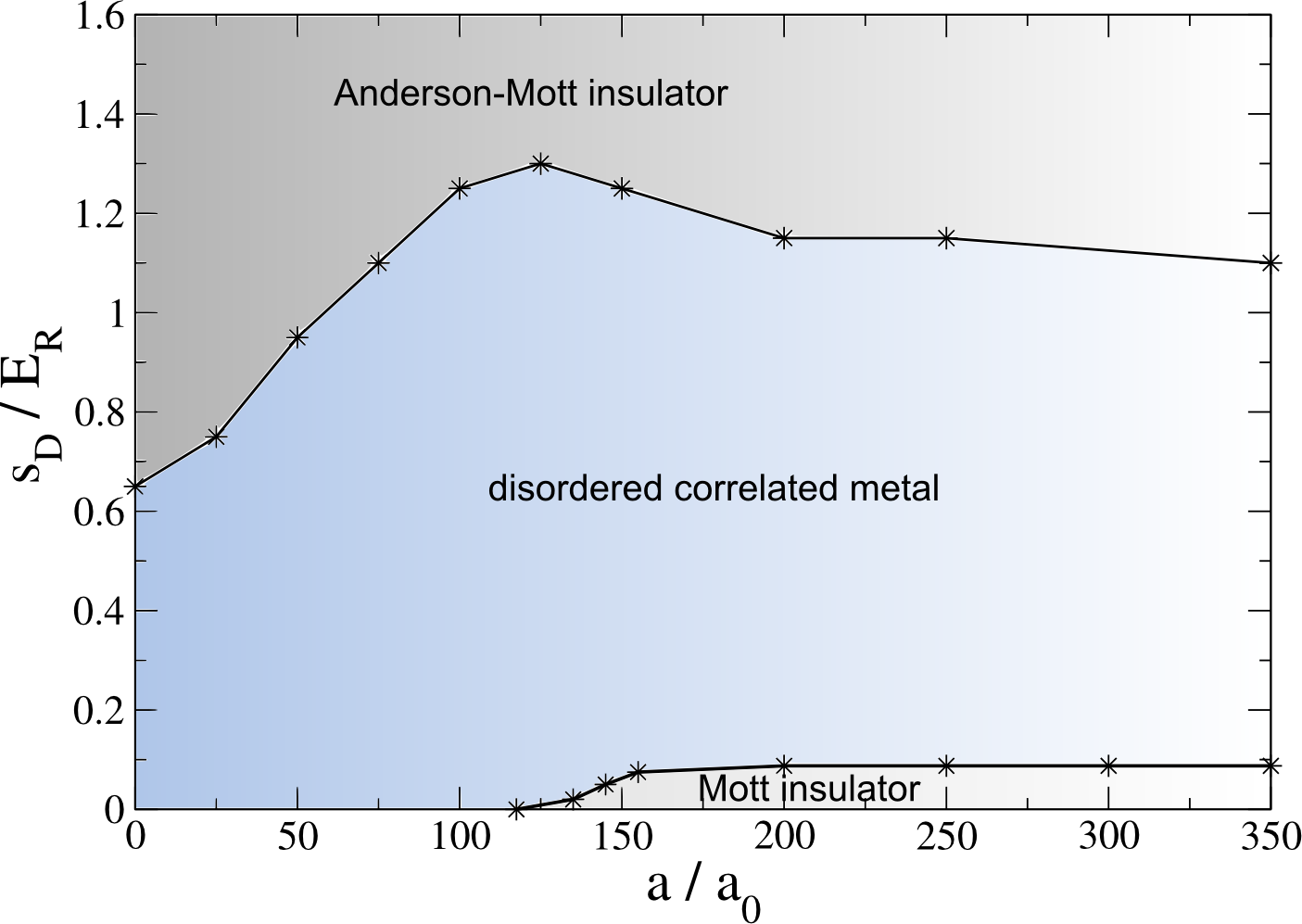}  
\caption{
(Color online) Paramagnetic ground state phase diagram of fermions in a speckle disordered optical lattice. Parameters are $\nu=1.0$ and $s_L = 10 E_R$. In the Munich experiment\cite{Schneider08} (so far without disorder) scattering lengths up to $a=300 a_0$ can be achieved.}
\label{figPD}
\end{figure}

Now we will discuss the main result of our investigation, i.e. the paramagnetic ground state phase diagram of interacting fermions exposed to speckle disorder (as a function of $s$-wave scattering length $a$ and speckle field strength $s_D$), as displayed in Fig. \ref{figPD}. 

In the absence of disorder ($s_D=0$), a Mott metal insulator transition is found at intermediate interaction strength. For the system considered here,  we found the critical $s$-wave scattering length for the Mott transition $a_c = 117.5 a_0$. In the absence of interactions ($a=0$), the Anderson transition occurs at $s_D=0.65$. In a system with both speckle disorder and interactions, three separate phases exist: Mott insulator, disordered correlated metal, and Anderson-Mott insulator. 

%%%%%%%%%%%%%%%%%%%%%%%%%%%%%%%%%%%%%%%%%%%%%%%%%%
%%%%%%%%%%%%%%%%%%%%%%%%%%%%%%%%%%%%%%%%%%%%%%%%%%%

\begin{figure}[b]
\includegraphics[width=0.47\textwidth]{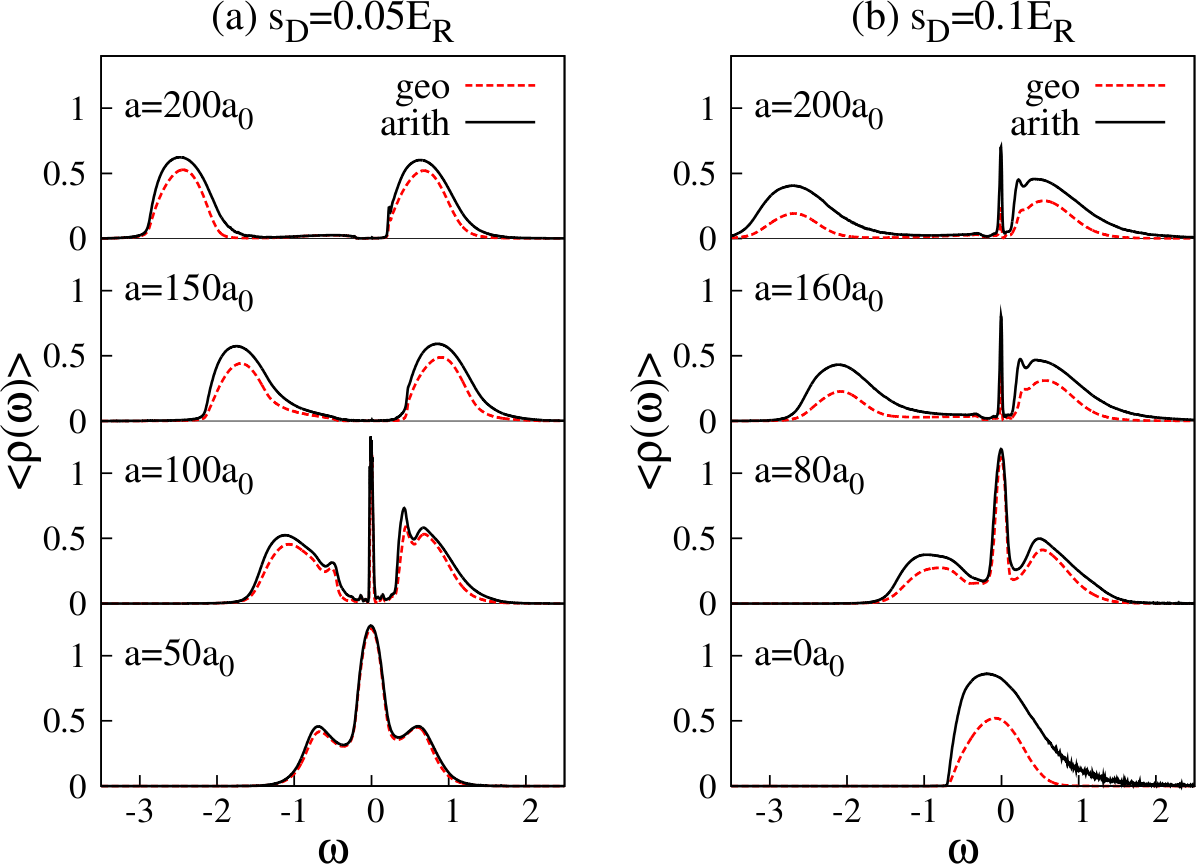}  
\caption{(Color online) Arithmetically (black solid) and geometrically (red dashed) averaged spectral function with increasing $s$-wave scattering length $a$ for speckle field strength $s_D=0.05 E_R$ in panel (a) and for speckle field strength $s_D=0.1 E_R$ in panel (b). Parameters are $\nu=1.0$ and $s_L = 10 E_R$.}
\label{fig_Mottsuppression}
\end{figure}

In Fig.~\ref{fig_Mottsuppression} the arithmetically  and geometrically averaged spectral functions are given for two different speckle field strengths $s_D$: in panel (a) for $s_D=0.05 E_R$ and in panel (b) for $s_D=0.1 E_R$. In both cases the interaction strength is increased from the bottom to the top.  Remarkably, the spectral functions evolve very differently for the two disorder strengths. For speckle field strength $s_D=0.05 E_R$ a correlation-induced metal-insulator transition takes place at finite $a$. For $s_D=0.1 E_R$ in the investigated regime up to $a=350 a_0$, no metal-insulator transition was found.  Instead, the Kondo peak, i.e. the coherent low-energy excitations, are stabilized for higher interaction values, whereas the lower Hubbard band is shifted away from the Fermi level. 

%%%%%%%%%%%%%%%%%%%%%%%%%%%%%%%%%%%%%%%%%%%%%%%%%%
%%%%%%%%%%%%%%%%%%%%%%%%%%%%%%%%%%%%%%%%%%%%%%%%%%%

This behavior is caused by the redistribution of states into the Mott-Hubbard gap due to disorder. For sufficiently strong disorder, the gap is closed. The unbounded nature of the speckle disorder at any finite $\Delta$ gives rise to states with very high energies, although their number is exponentially suppressed in $\Delta^{-1}$. This means that the Mott transition at finite disorder strength, which is described here,  might even be an artifact of the finite size $N$ in the stochastic Green's function ensemble. If this is the case, it would be an intrinsic feature of any finite size optical lattice as well. 

%%%%%%%%%%%%%%%%%%%%%%%%%%%%%%%%%%%%%%%%%%%%%%%%%%
\begin{figure}[tb]
\includegraphics[width=0.47\textwidth]{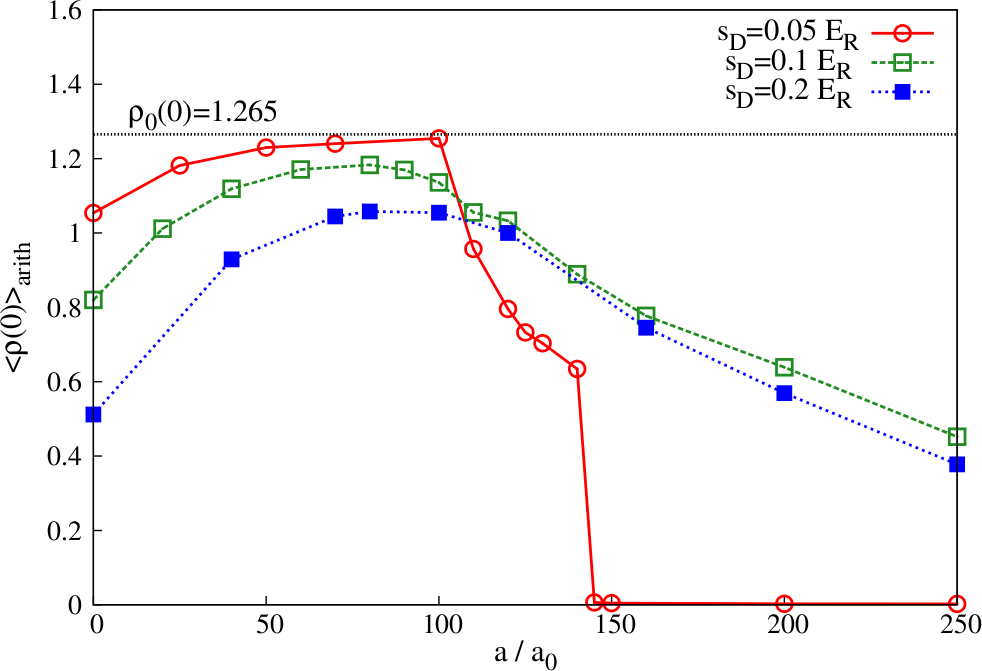}  
\caption{(Color online) Arithmetically averaged local density of states at the Fermi level $\langle\rho(0)\rangle_{\mbox{\tiny arith}}$ for three different values of the speckle field strength $s_D$: $s_D=0.05E_R$ (red solid line),  $s_D=0.2E_R$ (green dashed line) and  $s_D=0.2E_R$ (blue dotted line). For comparison the local density of states at the Fermi level of the homogeneous, non-interacting system $\rho_0(0)=1.265$ is included.  Parameters are $\nu=1.0$ and $s_L = 10 E_R$.}
\label{fig_rho_at_fl}
\end{figure}
%%%%%%%%%%%%%%%%%%%%%%%%%%%%%%%%%%%%%%%%%%%%%%%%%%%

To gain further insight, we plotted the arithmetically averaged LDOS at the Fermi level as a function of the $s$-wave scattering length $a$ in Fig.~\ref{fig_rho_at_fl} for three different speckle field strengths. Due to the finite disorder strength, the arithmetically averaged LDOS at zero interaction is reduced, which means that the Luttinger theorem is not fulfilled in presence of disorder. By increasing the interaction strength, the metallicity is improved for all three disorder strengths. At low disorder strength ($s_D=0.05E_R$), the Luttinger theorem is asymptotically fulfilled for $a\sim100a_0$. This is in agreement with results for the Anderson-Hubbard model with box disorder.\cite{Byczuk05,Byczuk10,Aguiar2009} The metallicity is suddenly reduced for stronger interactions and finally a Mott-Hubbard transition takes place at $a_c=145a_0$ for $s_D=0.05E_R$. For higher disorder strengths ($s_D=0.1E_R$ and $s_D=0.2E_R$), no Mott transition is found for scattering lengths up to $a=350a_0$.     
    
It is interesting to compare the qualitative structure of the phase diagram in Fig.~\ref{figPD} with the counterpart of ultracold bosonic atoms in speckle disordered lattices. In the latter case, an arbitrarily weak speckle field leads to a vanishing of the excitation gap and the Mott insulator only exists in the homogeneous system without disorder, in contrast to the fermionic case where a Mott insulator may exist at $\Delta>0$. Furthermore, the results presented here differ from the results obtained within typical medium theory\cite{Byczuk05,Byczuk10,Aguiar2009} for fermions with bounded box disorder. Although a delocalization tendency was found for box disorder, a correlation-induced metal-insulator transition takes place at intermediate disorder strengths. The critical interaction strength is shifted to higher values proportional to the disorder strength. Thus, the Anderson-Mott insulator and the Mott insulator were found to be continuously connected.\cite{Byczuk05} All important differences between the paramagnetic ground state phase diagrams for box disorder and speckle disorder can be attributed to the unbounded nature of the speckle distribution. In particular, the result that the Mott insulator and the Anderson-Mott insulator are not continuously connected is not attributed to the methodical differences between statistical DMFT and typical medium theory.

\subsection{Results within RDMFT}\label{rdmft}
%%%%%%%%%%%%%%%%%%%%%%%%%%%%%%%%%%%%%%%%%%%%%%%%%%
\begin{figure}[tb]
\includegraphics[width=0.48\textwidth]{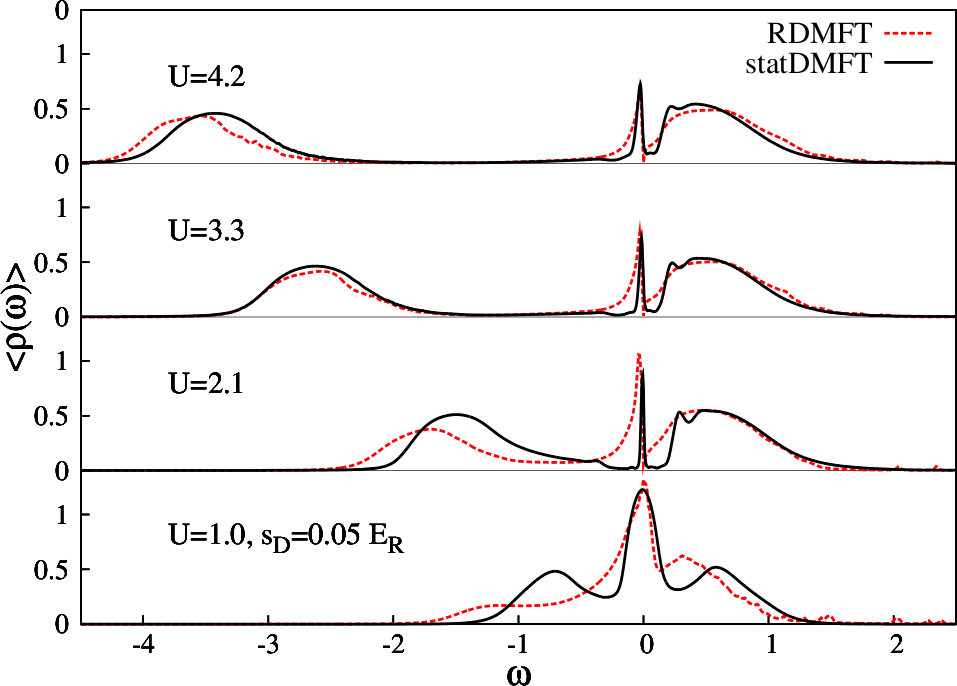}  
\caption{(Color online) Arithmetically averaged spectral function for increasing interaction $U$ at fixed speckle field strength $s_D=0.05E_R$ obtained via statistical DMFT with $K=4$ (black solid line) and real-space DMFT (red dashed line). The spin-summed filling is given by $\nu=1.0$ and the lattice size within the RDMFT calculations was $24\times24$.}
\label{fig_stat_RDMFT_comparison}
\end{figure}
%%%%%%%%%%%%%%%%%%%%%%%%%%%%%%%%%%%%%%%%%%%%%%%%%%%
In order to investigate the speckle disordered square lattice and to assess our results obtained within statistical DMFT for fermions with a semi-elliptical DOS, we performed complementary RDMFT calculations. An exemplary comparison of the arithmetically averaged spectral functions obtained by the two methods for identical parameters is given in Fig. \ref{fig_stat_RDMFT_comparison}. In these statistical DMFT calculations the connectivity $K=4$ is chosen to obtain the same bandwidth as on the square lattice. Throughout this section $U$ denotes the most probable value of the respective marginal PDF $p_U(U)\equiv \int d\epsilon \, p_{\epsilon,U}(\epsilon,U)$. 

Both methods lead to qualitatively identical results. The differences in the spectral functions can be traced back to the differences of the simulated models. The statistical DMFT is employed for particles with a semi-elliptical DOS, whereas RDMFT was applied for particles on the square lattice. Since the kinetic energy is connected to the lattice structure, the observed differences are pronounced when the kinetic energy dominates over the interaction energy. Consequently, deviations in the distributions of the spectral weights are larger for low and intermediate interaction strengths (Fig.~\ref{fig_stat_RDMFT_comparison} $U=1.0$ and $U=2.1$). On the other hand, the agreement is good for the strongly interacting case (Fig.~\ref{fig_stat_RDMFT_comparison} $U=3.3$ and $U=4.2$).

%%%%%%%%%%%%%%%%%%%%%%%%%%%%%%%%%%%%%%%%%%%%%%%%%%
\begin{figure}[b]
\includegraphics[width=0.48\textwidth]{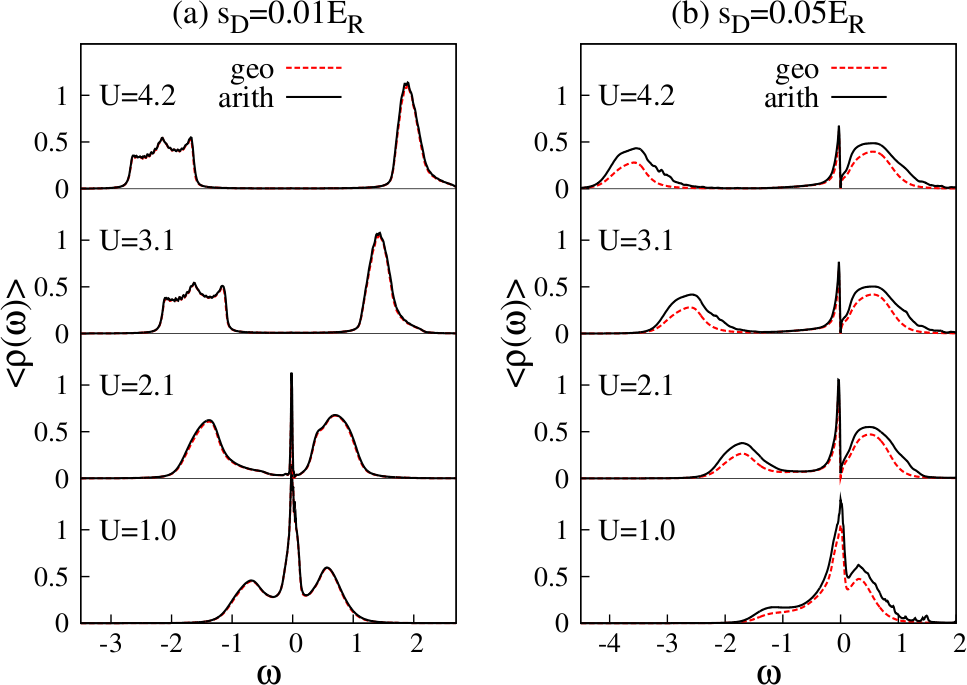}  
\caption{(Color online) Arithmetically (black solid) and geometrically (red dashed) averaged spectral function for increasing interaction strength $U$ at speckle field strength $s_D=0.01 E_R$ in panel (a) and $s_D=0.05 E_R$ in panel (b) obtained by RDMFT on a $24\times24$ square lattice. The spin-summed filling is given by $\nu=1.0$.}
\label{fig_rdmft_Mottsuppression}
\end{figure}
%%%%%%%%%%%%%%%%%%%%%%%%%%%%%%%%%%%%%%%%%%%%%%%%%%%

Interestingly, a pseudo-gap at the Fermi level in the LDOS on the square lattice is found within RDMFT. This pseudo-gap arises for intermediate and strong interactions in the presence of disorder and is stable under variation of the system size.  The pseudo-gap anomalies, also called zero bias anomalies, are a common feature in two-dimensional strongly correlated systems with disorder.\cite{Altshuler79,Efros75} A pseudo-gap anomaly was for instance found within a quantum Monte Carlo investigation of the Anderson-Hubbard Hamiltonian with box disorder.\cite{Chiesa08,Song09}

%\ref{fig_Mottsuppression}, 
Arithmetically and geometrically averaged spectral functions calculated by RDMFT for two different disorder strengths, namely $s_D=0.01E_R$ and $s_D=0.05E_R$, are displayed in Fig.~\ref{fig_rdmft_Mottsuppression}. Qualitatively, the spectral functions show similar behavior as obtained within statistical DMFT, cf. Fig.~\ref{fig_Mottsuppression}. For weak disorder ($s_D=0.01E_R$), metallic solutions are obtained for weak interactions. Raising the interaction, a Mott insulating phase is found, analogous to the case of a homogeneous system. On the contrary, for larger speckle disorder ($s_D=0.05E_R$) the LDOS remains finite at the Fermi level $\omega=0$, even at strong interaction $U=4.2$.

%%%%%%%%%%%%%%%%%%%%%%%%%%%%%%%%%%%%%%%%%%%%%%%%%%
\begin{figure}[tb]
\includegraphics[width=0.48\textwidth]{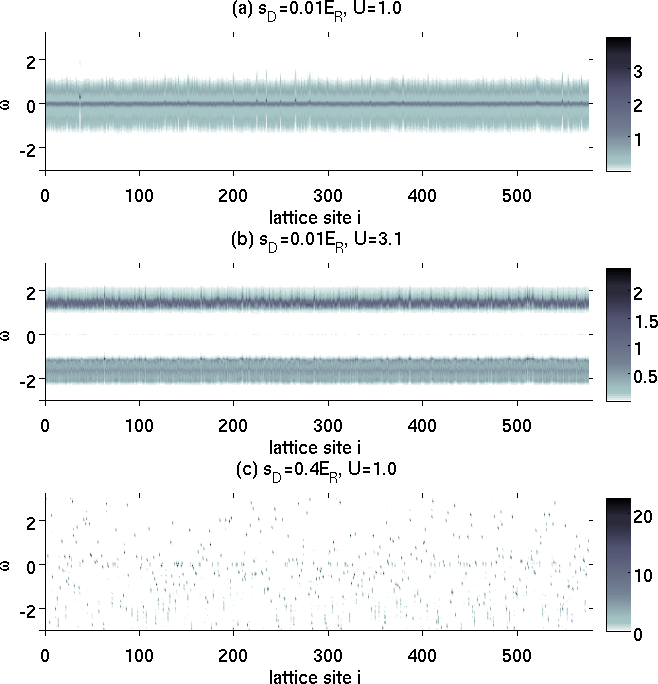}  
\caption{(Color online) Color coded local density of states $\rho_i(\omega)$ as a function of frequency $\omega$ and lattice site index $i$ for three different parameter sets: (a) $U=1.0$ and $s_D=0.01 E_R$, (b) $U=3.1$ and $s_D=0.01 E_R$ and (a) $U=1.0$ and $s_D=0.04 E_R$. Parameters are $\nu=1.0$ and the lattice size was $24\times24$.}
\label{fig_RDMFT_color}
\end{figure}
%%%%%%%%%%%%%%%%%%%%%%%%%%%%%%%%%%%%%%%%%%%%%%%%%%%

We note that within RDMFT for  $U=3.1$ and $U=4.2$, the lower Hubbard-band shows a peaked structure at low disorder. This feature cannot be exclusively identified with physical properties of the system because of numerical uncertainties. Within RDMFT, an artificial broadening $\eta$ is applied, which is scaled proportionally to the system size, i.e. $\eta \propto 1/L^2$. Since the peaks are not fully recovered for other lattice sizes, we conclude that they are finite size effects.

The real space resolution of the LDOS gives us insight into localization effects of the system. In Fig.~\ref{fig_RDMFT_color} the LDOS $\rho_i(\omega)$ is plotted for a $24\times24$ lattice and different interaction strengths, each for a different disorder realization. At $U=1.0$ and weak disorder $s_D=0.01E_R$ (Fig.~\ref{fig_RDMFT_color} panel (a)) the spectral weight around the Fermi level $\omega=0$ remains finite at each lattice site. The vast majority of single particle states are extended and the system is in the metallic phase. At $U=3.1$ (Fig.~\ref{fig_RDMFT_color} panel (b)) Hubbard bands are formed and the spectrum exhibits a gap proportional to the interaction strength for all lattice sites, indicating that the system is in a Mott insulating state. However, as the speckle field strength is increased to $s_D=0.4E_R$, the states with spectral weight at the same frequency are spatially separated. In other words, the spectrum is highly fragmented and each local spectrum consists of isolated delta peaks, consistent with Anderson localized states in the infinitely large system (cf. Fig.~\ref{fig_RDMFT_color} panel (c)).

\subsection{Finite temperature}\label{sec_finiteT}

\begin{figure}[tb]
\includegraphics[width=0.47\textwidth]{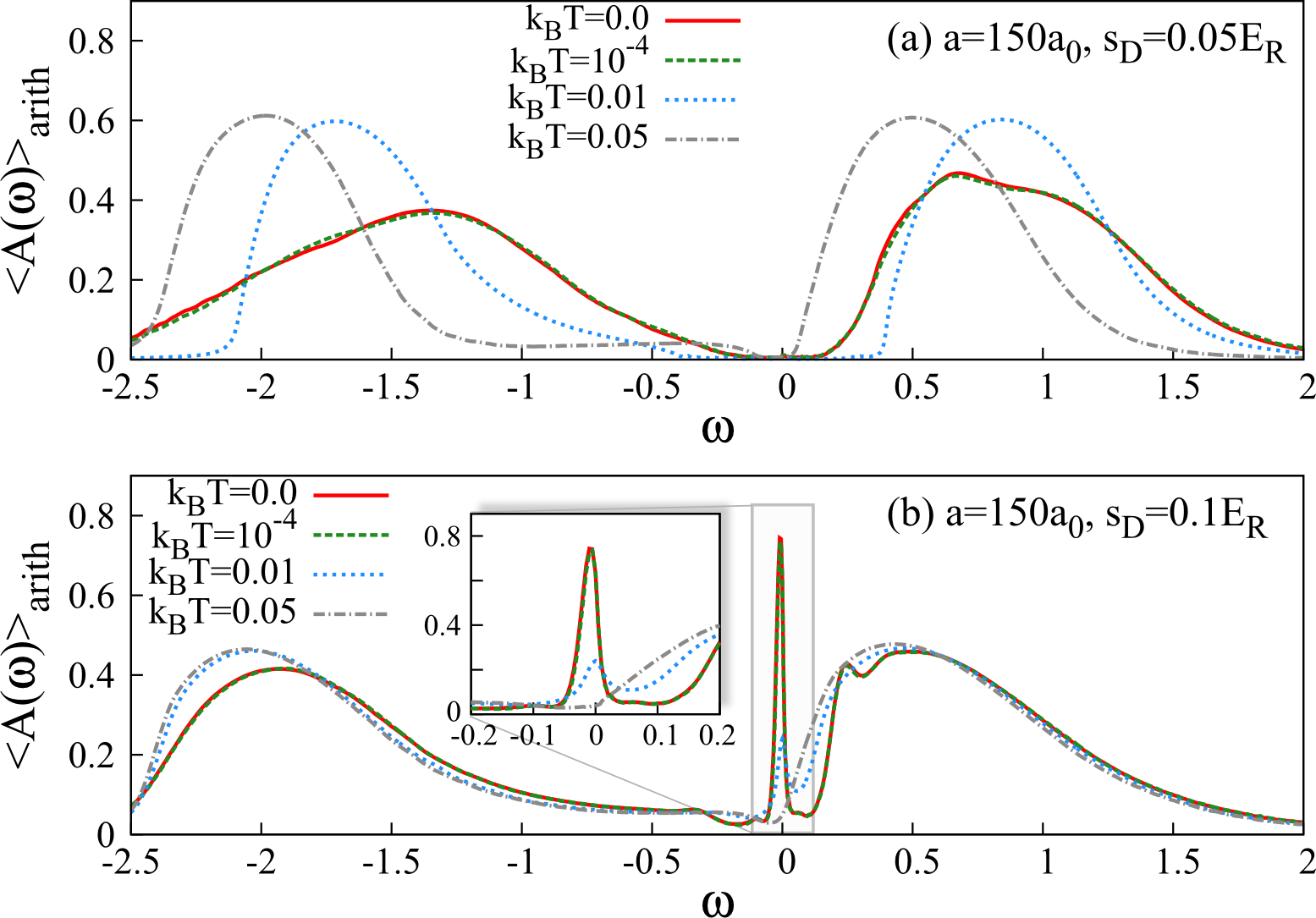}  
\caption{(Color online) Evolution of the arithmetically averaged spectral function with increasing temperature for two parameter sets: (a) $s_D=0.05 E_R$, $a=150 a_0$ and (b)  $s_D=0.1 E_R$, $a=150 a_0$. Parameters are $\nu=1.0$ and $s_L = 10 E_R$.}
\label{fig_fT1}
\end{figure}

Here, we investigated a system of fermions with a semi-elliptic DOS at finite temperature. The spectral functions are plotted in Fig.~\ref{fig_fT1} for two parameter sets. Panel (a) displays the arithmetically averaged spectral functions for speckle disorder strength $s_D=0.05$ and scattering length $a=150a_0$ for various increasing temperatures. At zero temperature, this parameter set would correspond to the Mott insulator. In Fig. \ref{fig_fT1} we note that with increasing temperature, the gap initially grows and the incoherent excitations reveal a significant redistribution of the spectral weight, which is shifted away from the Fermi level. This is not observed in the homogeneous case, where the spectral transfer is weak. Panel (b) shows the corresponding behavior for stronger disorder, namely $s_D=0.1$ and $a=150a_0$, which corresponds to the disordered strongly correlated metal at zero temperature. With increasing temperature, the coherent low energy excitations are reduced, and for $k_BT =0.05$ the system enters a Mott insulating state. 

%%%%%%%%%%%%%%%%%%%%%%%%%%%%%%%%%%%%%%%%%%%%%%%%%%
%%%%%%%%%%%%%%%%%%%%%%%%%%%%%%%%%%%%%%%%%%%%%%%%%%%

\begin{figure}[tb]
\includegraphics[width=0.47\textwidth]{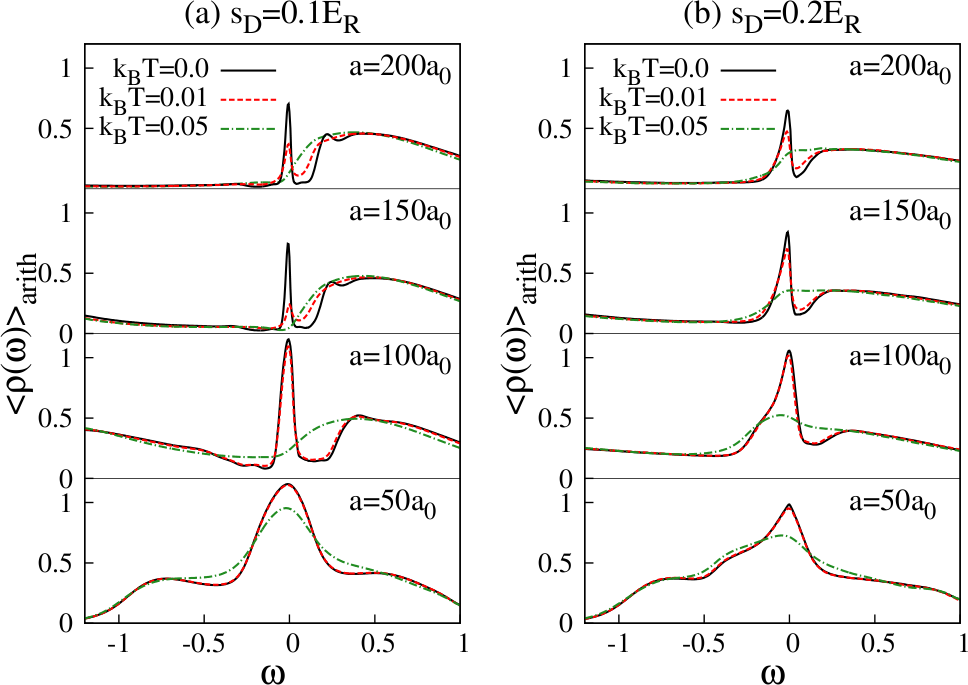}  
\caption{(Color online) Evolution of the arithmetically averaged spectral function with increasing $s$-wave scattering length ($a=50a_0,100a_0,150a_0,200a_0$) for two disorder strengths (a) $s_D=0.1E_R$ and (b)  $s_D=0.2 E_R$. For each parameter set the spectral function is compared at three different temperatures, namely $k_bT=0$ (black solid line), $k_bT=0.01$ (red dashed line) and $k_bT=0.05$ (green dash-dotted line). Parameters are $\nu=1.0$ and $s_L = 10 E_R$.}
\label{fig_fT2}
\end{figure}

In Fig. \ref{fig_fT2}, the evolution of the arithmetically averaged spectral function with increasing temperature is displayed for two values of the speckle field strength ($s_D=0.1E_R$ in panel (a) and $s_D=0.2E_R$ in panel (b)) and three different temperatures: $k_BT=0$, $k_BT=0.01$ and  $k_BT=0.05$. We note the reduction of the spectral weight at the Fermi level due to finite temperatures. This leads to an enlargement of the Mott insulating phase. However, as can be seen in panel (b) of Fig. \ref{fig_fT2}, a metal without a resonant peak at the Fermi level is stabilized a higher disorder strength. In this respect, our central finding at $T=0$, that the Mott and the Anderson-Mott insulators are not continuously connected in presence of the speckle disorder, also holds at finite temperature.

\section{Summary and outlook}\label{summary}

In conclusion, we have investigated a gas of ultracold fermions in an optical lattices subjected to an additional speckle disorder field using statistical DMFT and RDMFT. The presented DMFT schemes include off-diagonal hopping disorder and allow for a systematic inclusion of correlations between the difference in neighboring on-site energies and the hopping amplitude, as well as between the on-site energy and the local interaction strength, which are non-trivial in realistic experiments and have been calculated by Zhou and Ceperley.\cite{Zhou10}
%%%%%%%%%%%%%%%%%%%%%%%%%%%%%%%%%%%%%%%%%%%%%%%%%%
%%%%%%%%%%%%%%%%%%%%%%%%%%%%%%%%%%%%%%%%%%%%%%%%%%%

The complete paramagnetic ground state phase diagram has been determined. It consists of a disordered metallic phase, as well as Mott insulating and Anderson-Mott insulating phases. A strong suppression of the correlation-induced metal insulator-transition by disorder is observed and a finite metallic phase is found, even in the strongly interacting regime. Hence, the Mott and Anderson-Mott insulators are not continuously connected, in contrast to the prediction by typical medium theory in the presence of bounded box disorder. 
 
We have also investigated speckle-disordered fermions on a square lattice by means of RDMFT. We find that our main results for the case of high-dimensional systems also hold in two spatial dimensions. Moreover, a pseudo-gap was found, which should be investigated in detail in the future. Finally, we investigated the high-dimensional system at finite temperature, where the Mott insulating region is enlarged, but the separation of the Mott and the Anderson-Mott insulators persists.

\section*{Acknowledgments}

We are grateful to S. Q. Zhou and D. M. Ceperley for sharing the data of their calculations. Moreover, we acknowledge useful discussions with B. DeMarco, U. Schneider and I. Titvinidze. This work was supported by the Deutsche Forschungsgemeinschaft (DFG) via Forschergruppe FOR 801. Computations were performed at the Center for Scientific Computing (CSC) at the Goethe University Frankfurt. K.B. acknowledges support by the grant TRR80 of the Deutsche Forschungsgemeinschaft and the grant No. N~N202~103138 by the Polish Ministry of Science and Education.

\appendix

\section{Sampling the Hubbard parameter distributions}\label{appendix}

The various Hubbard parameters in an optical lattice exposed to speckle disorder with a realistic autocorrelation length of $~1\mu m$, such as realized in the experiments in the DeMarco group for instance,\cite{White09} were calculated by S. Q. Zhou and D. M. Ceperley\cite{White09,Zhou10} at fixed disorder strength $s_D=1E_R$ and lattice intensity $s_L=14 E_R$. Imaginary time evolution was used, calculating the parameters $U$, $\epsilon$ and $t$ for $1222$ disorder realizations on a three-dimensional $6\times 6 \times 6$ lattice. Based on these data, PDFs were constructed, which are sampled and scaled appropriately in our calculations for the various parameter regimes of interest.

This scaling behavior will shortly be elucidated in the following: 
The most likely value and lower bound for the on-site energy corresponds to regions of low intensity, where we set $\epsilon=0$, thus recovering the usual energy reference point in the absence of the disorder speckle laser. It is found in Ref.~(\onlinecite{White09}) that the disorder strength $\Delta$ (i.e. the standard deviation of $p_{\epsilon}(\epsilon)$) scales linearly with the speckle intensity $s_D$, as is to be expected in the perturbative limit. Thus, after an on-site energy is sampled from the distribution, it is simply scaled by multiplication with the respective $s_D$. An affine shift is not required due to the vanishing lower bound discussed above.

Since the on-site energy $\epsilon$ and interaction parameter $U$ are correlated, as can be seen from the histogram Fig.~\ref{fig_pd_e_U} (i.e. this two-dimensional PDF is not reconstructible as a product of two one-dimensional PDFs), a combined sampling from a two-dimensional distribution is used for $\epsilon$ and $U$. This is equivalent to first sampling $\epsilon$ from the marginal distribution $p_{\epsilon}(\epsilon)$ and subsequently $U$ from the conditional PDF $p_U(U|\epsilon)$. Furthermore, the standard deviation of the on-site interaction parameter $U$ (i.e. of the marginal distribution $p_U(U)\equiv \int d\epsilon \, p_{\epsilon,U}(\epsilon,U)$) scales linearly with $s_D$, while the most probable value of $U$ remains unaffected by a variation in $s_D$ up to a good approximation. Therefore, the variation in $U$ in the two-dimensional distribution $p_{\epsilon,U}(\epsilon,U)$ is scaled by $s_D$, while the most likely value is set to the value of $U$ determined from a band structure calculation of the Wannier functions in the pure case without disorder, as performed in Ref.~(\onlinecite{Greiner2003}) for instance.

In contrast to the local interaction parameter $U$, the nearest neighbor tunneling amplitude $t_{i,j}$ exhibits only very weak correlation with the respective on-site energy, while the correlation with the energy difference between the two sites $\Delta \epsilon\equiv \epsilon_i-\epsilon_j$ is significant. Hence, the distribution for $\Delta \epsilon$ and $t$ cannot be sampled independently (i.e. $p_{\Delta \epsilon, t}(\Delta \epsilon, t)\neq p_{\Delta \epsilon}(\Delta \epsilon) \, p_t(t)$) and a conditional distribution function $p_t(t|\Delta \epsilon)$ for $t$, given a fixed value of $\Delta \epsilon$, was constructed from the data in \cite{Zhou10}: For $200$ discrete values of $\Delta \epsilon$ a histogram approximating $p_t(t|\Delta \epsilon)$ was extracted, approximating the PDF. This PDF is integrated with respect to $t$, yielding the conditional PDF $F_t(t|\Delta \epsilon)$ and subsequently normalized for each fixed $\Delta \epsilon$, such that $\lim_{t \to \infty} F_t(t|\Delta \epsilon)=1$. To randomly sample values in a numerically efficient manner from $p_t(t|\Delta \epsilon)$, the conditional cumulative PDF $F_t(t|\Delta \epsilon)$ is inverted with respect to $t$ on a linearly interpolated grid on $[0,1]$, consisting of $800$ points. Given a fixed $\Delta \epsilon$, a randomly drawn value of the inverted cumulative distribution is thus distributed according to the initial conditional PDF $p_t(t|\Delta \epsilon)$, leading to the sought-after sampling routine.

\end{document}